\documentclass[prb,preprint,epsfig,floats]{revtex4}
\usepackage{graphicx,epsf}

\begin{document}


\title{ Thirty Years of heavy Fermions: Scientific Setting for their Discovery and Partial Understanding}
\author{C. M. Varma}
\address{Physics Department, University of California, 
Riverside, CA 92507}
\begin{abstract}
This is the text of a one of the talks given at the plenary symposium entitled "Thirty years of Heavy Fermions" at the beginning of the International conference on Strongly correlated electrons in Vienna in July 2005. Heavy-Fermions provide an extreme example of the utility of the idea of continuity and analyticity in physics. Their discovery and study in the past thirty years has added a fascinating chapter to condensed matter physics. I briefly review the origins of the heavy-fermion problem out of the study of magnetic moments in metals and the study of mixed-valent rare-earth compounds. I also review the principal ideas underlying the features understood in their fermi-liquid phase as well as in their anisotropic superconductivity. The unsolved issues are also briefly mentioned.
\end{abstract}
\maketitle
\section{Early History}
One of the major activities of the condensed matter physics community in the last 30 years has been the study of heavy fermions and heavy fermion superconductivity.
In this talk I wish to recount to you the scientific setting in which the heavy fermions and heavy fermion superconductivity were discovered and how they were at least partially understood. I will try to convey why I think they were so exciting to study and why the excitement still persists.

In 1975/1976, Hans Ott came as a visitor to Bell labs from ETH-Zurich to study the thermodynamic and transport properties of $CeAl_3$ to work with Klaus Andres, also an alumni of ETH and John Graebner \cite{ago75}.  They had decided to study this compound because several $Ce$ metallic compounds (and metallic $Ce$ itself) were known which did not order magnetically and had specific heats $\propto \gamma T$ at low temperatures with high values of $\gamma$, which could be expressed as a large effective mass 
$m^*/m$ of $O(10^2)$, and similarly enhanced magnetic susceptibility. Other rare-earth metallic compounds with similar properties contained $Yb$ or $Sm$. From lattice constant studies and occasionally through Mossbauer and core-level Photoemission studies, it was beginning to be understood that in these compounds the rare-earth valence fluctuated quantum-mechanically compared to the usual situation of stable valence. Yakov Yafet and I \cite{vy76} had just formulated  the problem of the fluctuating-valence or mixed-valence compounds at the time of Hans Ott's visit.
The stable valence and concomitant local magnetic moments which order magnetically at low temperatures, which is the state for most rare-earth compounds, were already well understood as due to the large local repulsion and  the weak hybridization of the $4f$-shell and the RKKY interaction between moments in metals. These followed from the work of Friedel and Blandin \cite{blandin-friedel} and Anderson \cite{anderson-localmoments} who had derived the conditions for formation of magnetic moments by impurities in metals. 

The interest in rare-earth compounds followed the study of the effect of magnetic impurities in superconductors, which was most prominently carried out by Berndt Matthias, Brian Maple, Dieter Wohleben, Brian Coles, Ted Geballe, Tournier, Tholence and others \cite{maple-radosuhl},\cite{wohleben-radosuhl}. It was observed that Ce
impurities behaved rather peculiarly and this was correctly attributed to the Kondo effect which was beginning to be understood in the late sixties and early seventies.

This work was of-course influenced by the theory of the effect of time-reversal violation  in superconductors due to magnetic impurities but the experiments had important influence on the theorists as well. Anderson relates in his 1967 Les Houches lectures that the issue was framed to him in a typically succinct fashion by Matthias: " A few percent iron in Pb reduces $T_c$ to $0$. You tell me that this is due to the magnetic moment of iron impurities. But the same iron increases the $T_c$ in $Mo$. What do you mean that iron impurities are magnetic?". 
 
The most prominent intellectual antecedent of the heavy -fermion and the mixed-valence problems is of-course the discovery of the Kondo problem in 1964 \cite{kondo} and
its solution by Wilson \cite{wilson} and by Yuval and Anderson \cite{yuval-anderson}, followed by others. The most useful manner of thinking about the Kondo problem from point of view of heavy-fermions is the  formulation by Nozieres \cite{nozieres}, who re-expressed the solution of the problem in the language of a (local) fermi-liquid theory.

Yafet and I were studying these problems in the stimulating intellectual atmosphere of Bell labs where the Kondo problem was actively discussed and many experiments on related problems were being done and where we had many visitors discussing all the important developments in every part of the world. It was easy to see that the mixed-valence compounds had avoided local magnetism of the rare-earths through the hybridization process: $f^{n} \to f^{n-1}(ds)^1$, i.e. by the hybridization of $f-$orbitals on an atom with the $s-d$ orbitals of the neighbors with the same symmetry as the $f$. This hybridization had to be the most important process in destabilizing the rare-earth moments  because of the near degeneracy of these configurations in $Ce, Sm$ and $Yb$, due to simple atomic physics. But the relation of this hybridization with the Kondo effect was quite unclear. So we first looked at the single impurity problem with a variational method and showed that if the hartree-fock resonance of the local orbital $e_f$ measured with respect to the chemical potential was much farther than the hybridization width $\Gamma$, the energy of the ground state singlet was lower than that of the doublet or the triplet by the customary expression for the Kondo temperature $T_K$ but in the opposite limit the singlet ground state was bound by the $O(\Gamma)$. There was therefore a clear relationship between the degree to which the hartree-fock resonance is unoccupied and the Kondo temperature which could be expressed as \cite{vy76} 
\begin{equation}
T_K \approx \Gamma (1-<n_f>). 
\end{equation}
In the same paper we introduced the model for the lattice which came to be known as the Anderson lattice model and solved it in the simplest approximation due to Hubbard of de-coupling the constrained kinetic energy operators. This led us to the hybridized band picture Fig.(1), where the band-width of the narrow band is $O(T_K)$ defined as in Eq.(1). For the mixed-valence compounds,
$(1-<n_f>)$ is about 0.5. This led to the right magnitude of specific heat and magnetic susceptibility enhancements of $O(10^2)$.

\begin{figure}[!ht]
\begin{center}
\includegraphics[width=0.6\textwidth]{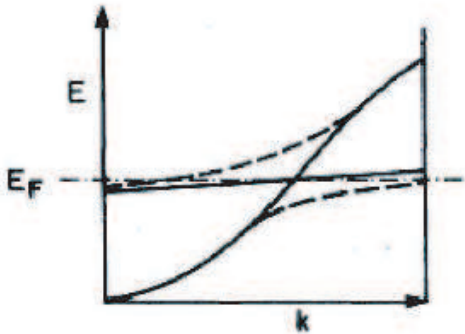}
\end{center}
\caption{Band-structure of mixed-valence and heavy-fermion compounds, taken from C.M. Varma Rev. Mod. Phys. {\bf 48}, 219 (1976). }
\label{fig1}
\end{figure}

Hans Ott told me at the time of their results  that the efffective mass enhancement in their measurement in $CeAl_3$ was more than an order of magnitude larger than this and asked me what I thought may be different from the other $Ce$ compounds. Having the results Eq.(1), I was able to tell him that in $CeAl_3$ ,due to accidents of chemistry,  the f-level must be almost integer valent, i.e. $(1-<n_f>) \lesssim 0.1$, i.e. the tail of the f-hartree-fock resonance intersects the chemical potential. This suggestion with due reference was noted in the Andres-Graebner-Ott paper \cite{ago75}.

With $CeAl_3$, the era of heavy-fermi liquids started. It was fascinating that the entropic (and the dynamic) mass of an electron in a solid due to many body effects could be similar to that of a proton and yet it could be thought of at low enough temperatures in the same way conceptually as an electron in metallic sodium. This is the most extreme example known of the power of the analyticity and continuity arguments of Landau. In Cuprates as indeed in heavy-fermions near their quantum-critical point, the analyticity argument breaks down and we have come to appreciate the difference between effective mass enhancements of several thousand and a mass $\propto \log(\omega_c/T)$ even in the range of $T$ where this quantity is only about 3. This is because the scattering rate in fermi-liquids is $\propto T^2$, while in the Cuprates it is
$\propto T$, as follows from the Kamers-Kronig transform between the real and imaginary part of the self-energy.

\section{Further Developments}

Eq.(1) and indeed the whole story above has been derived in a number of new and increasingly sophisticated methods: the non-crossing approximation \cite{keiter-grewe}, the slave-Boson approach \cite{coleman}, the large N (degeneracy) approximation \cite{rama-sur}, the Gutzwiller method \cite{rice-ueda} and other methods \cite{fulde-razah}.  Eq.(1) is found to be exact only in the large N limit. To my mind all these methods are essentially equivalent. Some are of-course more beautiful mathematically, like for example the slave Boson methods. Their seductive nature in turn  has led to much abuse, especially in their use in the Cuprate problems. The real next theoretical advance can come about by the appropriate use of the cluster extensions of the DMFT \cite{dmft} methods, which incorporate the incoherent parts of the bare-particles in the calculations as well as the effects of inter-site interactions, which are neglected by the other methods. 

There were several interesting  issues raised by the mixed-valence and heavy fermion compounds. Where is the large entropy and spin-susceptibility coming from?  This was understood by comparing the entropy vs. temperature of the heavy-fermion compounds with that of the typical rare-earth magnetically ordered compound \cite{cmv-prl85}. It is clear from fig.(2) that in heavy-fermions the spin-entropy is converted to a fermi-liquid entropy just as in the Kondo effect. 

 \begin{figure}[htbp]
\begin{center}
\includegraphics[width=0.5\textwidth]{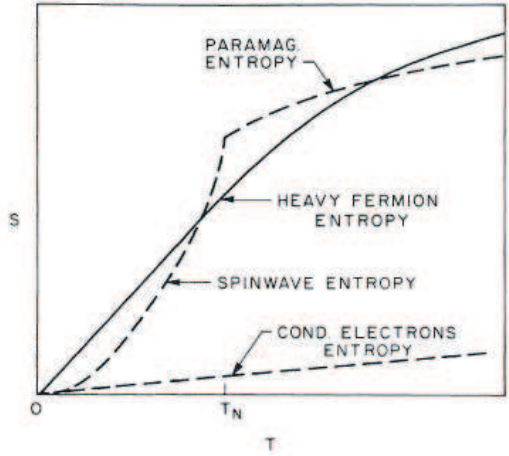}
\caption{Using Entropy conservation argument between a rare-earth compound which orders magnetically below $T_N$ and a heavy-fermion compound to show that the heavy-fermi-liquid entropy comes from converting spins to fermions. The dashed lines refer to the entropy contribtion of the former and the solid line to the entropy of the latter. Taken from Ref.(17).}
\label{fig. 2}
\end{center}
\end{figure}

\subsection{Fermi-liquid Parameters}

The next question was if these compounds are fermi-liquids, what are the Landau parameters. An important hint, as usual, was given by the experiments.  Bertram Batlogg was studying the temperature dependence ultrasound attenuation in $UPt_3$ together with David Bishop \cite{bishop} to test if it was indeed an anisotropic superconductor. Bertram told me that there is a surprise already in the normal state. The magnitude of the ultrasound attenuation was the same as in an ordinary metal such as $Sn$. The ultrasound attenuation at low temperatures is
proportional to 
\begin{equation}
(m^*/m)^2 <|M_{def}|^2>
\end{equation}
Here $<|M_{def}|^2>$ is the square of the matrix element of the deformation potential for the fermions between low energy states. Since $m^*/m$ is enhanced by a factor of $O(10^3)$, one could only understand the experimental result  if the fermi-liquid renormalizations are such that $<|M_{def}|^2>$ are renormalized downwards (Vertex Correction! ) by the same factor. This matrix element has four powers of wave-functions each of which has a quasi-particle amplitude $z$. So each $z$ had to be renormalized  downwards  by  $(m/m^*)^{1/2}$. This sort of thing does not happen, for instance, in liquid $^3He$. 
At the same time Shelly Schultz at LaJolla was doing EPR experiments on impurities  in $UBe_{13}$ \cite{gandra} told me that there are no signs of the enhanced $m^*/m$ in his measurements as opposed to the NMR measurements of Maclaughlin on $Be$ in $UBe_{13}$ \cite{maclaughlin} . Again the matrix elements had to cancel the enhancement of $m^*/m$ in the EPR experiments, but the important matrix elements for $Be$ nuclei are the direct dipolar interactions which do not involve  the heavy fermion wavefunctions and are not renormalized. The thermal conductivity in heavy-fermions also does not show any signs of renormalization \cite{floquet}.

It turns out that the above cancellations, which can be expressed as the cancellation of renormalization in $\tau^*/m^*$, where $\tau^*$ is the heavy-fermion relaxation rate, come about  if the one particle self-energy of the fermions $\Sigma({\bf k},\omega)$ is independent of $v_f{\bf k}$ compared to its dependence on $\omega$ \cite{cmv-prl85}. This happens for instance in electron-phonon interactions \cite{prange}. But as opposed to electron-phonon interactions, the entity scattering had to be spin-fluctuations on a scale of $T_K$ for essentially all ${\bf k}$, so that the magnetic susceptibility is enhanced on the same scale as the specific heat. One of the consequences of the ${\bf k}$ independence of $\Sigma$ is that the specific heat enhancement is given simply by $z^{-1}$, where $z$ is the quasiparticle renormalization.
This is indeed very different from the translationally invariant case of liquid $^3He$, where $z$ must drop out of all low-energy properties. The picture that this renormalization provides of many body effects due to exchange of spin-fluctuations with a Migdal like theorem operating makes the heavy-fermion problem, a systematically calculable problem in the Fermi-iquid regime for $T << T_K$. Empirical relations such as due to Kadowaki-Woods \cite{kadawaki} between the resistivity $\rho(T)\propto AT^2; A \propto (m*/m)^2$ can for instance be proved easily \cite{miyake-matsu-cmv}. At the same time, the resistivity due to impurities is not renormalized. Following Prange and Kadanoff \cite{prange}, for the case that the renormalizations are due to exchange of spin-fluctuations and self-energy has only negligible ${\bf k}$-dependence, the following relations hold for heavy-fermions : 
\begin{eqnarray}
m^*/m =1/z=(1+F_0^s); \\ \nonumber
 \chi/\chi_0 = (m^*/m)/(1+F_0^a),
\end{eqnarray}
and the compressibility is not renormalized. 
The dynamic mass, i.e. the inertial mass in response to electromagnetic fields is the same \cite{vms} as the entropic or specific heat mass $m^*$, again different from a Galilean fermi-liquid like $^3He$.

The most complete verification of the independence of $\Sigma({\bf k},\omega)$ on ${\bf k}$ came through the Fermi-surface measurements of the Cambridge group \cite{cambridge}. The measured fermi-surface agrees very well with that calculated by ordinary band-structure methods although the effective mass is two orders of magnitude larger than such calculations.

In relation of the heavy-fermions to the Kondo effect, one had to understand why the resistivity in the former increases as $T^2$, while in the latter it decreases as $T^2$.
The point is that in the Kondo effect the phase-shift as a function of energy $\omega$ (effectively as a function of temperature) can be divided up into an elastic part which decreases as $\omega^2$ from the unitarity limit of $\pi$ at $T=0$ and an inelastic part which increases as $\omega^2$. In a periodic lattice the elastic part of the phase-shift is simply absorbed as Umklapp scattering leading to a band-structure. So only the inelastic part influences the transport as a function of frequency or temperature.

\subsection{Interactions between Moments}

The heavy Fermions posed a few problems not encountered in Mixed-valence compounds. Almost no mixed-valence compounds have magnetic order or superconductivity unlike the heavy Fermions. In mixed-valence compounds the dominant magnetic interaction between local moments is double-exchange rather than the RKKY interaction since there can be no RKKY interaction if the charge fluctuation energy is much larger than the RKKY interaction energy estimated with the assumption of no charge fluctuations. Moreover in compounds of $Ce, Yb$ and $Sm$, there is no double-exchange either because one of the valence states is non-magnetic. (Only mixed-valence compounds of $Tm$ have both valences magnetic; these were predicted and found to have a ferromagnetic transition).  
A qualitative answer that the interactions do not matter if the Kondo temperature is larger than the interaction energy which was offered by Doniach \cite{doniach} and by me\cite{vy76} is not really satisfactory because for any spin-1/2 problem, it is impossible to achieve this condition. The condition is achieved for large enough degeneracy N because  
$T_K \propto \exp(1/NJ\rho)$ while $I \propto NJ^2/\rho$. But the ground state of many heavy fermion compounds is a doublet $(N=2)$. The answer to this dilemma has probably to do with the modification of the Kondo effect and RKKY interactions due to higher lying crystal field levels. This leads to a crossover between different Kondo temperature scales which are in evidence in several compounds.

The regime of $I$ and $T_K$ of similar magnitude is achieved in many heavy-fermion compounds and is actually the most active current problem in the field as it leads to the competititon between heavy-fermion behavior, anisotropic superconductivity and magnetic order. It appeared interesting to work out the detailed theory of two interacting Kondo impurities. This was first done with the Wilson RG technique \cite{jones-varma} and subsequently by analytic methods \cite{affleck-ludwig}. While attention has focussed on the quantum critical point in the problem, the more interesting physics for understanding the heavy fermions seem to me to be the slow crossover between the  fixed point in which a singlet state  forms due to  the AFM-interactions between the  moments the singlet state due to the Kondo effect at each site. Also interesting is the low temperature effective hamiltonian \cite{jones-varma} which gives the low energy spectra. This exhibits pairing interactions as well as magnetic ordering interactions of the quasi-particles. 

\section{Heavy-Fermion Superconductivity}

In 1979 Frank Steglich and collaborators discovered \cite{steglich} superconducitivity in the heavy fermion compound $CeCu_2Si_2$
and there ensued a
 period in which the idea that  this could not be conventional phonon induced superconductivity met a lot of resistance. The physics community is conservative and resistant to new things. This can be frustrating but is on the whole a good thing because new things ought to be accepted only after they are further corroborated.  In a few years Fisk, Smith and Ott discovered \cite{fisk-smith-ott} superconductivity in $UBe_{13}$ and Stewart and collaborators \cite{stewart} in $UPt_3$ and now there are probably more than 20 heavy Fermion superconducting compounds. This has been accompanied by a systematic calculation of the origin of superconductivity and calculation of properties and experiments proving the anisotropic spin-fluctuation induced superconductivity. Indeed these ideas are now so well accepted that they have become part of the orthodoxy along with phonon induced superconductivity. Inevitably, they have been misapplied to the next great thing that came along - the Cuprate superconductors.

It was evident \cite{cmv-baps} for two reasons that heavy-fermion superconductivity could not be due to phonons; (1) Everything that one has learnt from the microscopic theory of pairing (BCS and Eliashberg theories) tells us that superconductivity is caused by the same excitations which dominate the scattering of fermions in the normal state to determine properties like the resistivity. In heavy fermions, the inelastic part of the resistivity is proportional to $T^2$; therefore the dominant scattering is fermion-fermion scattering. (2) In most of the heavy-fermion superconductors, the fermi-energy is smaller than the typical phonon energies. In that case the phonon interactions are not retarded in relation to the fermion-fermion interactions and hence there is no net attraction integrated over the characterisitic range of fermion energies. 

Hindered by the knowledge about $^3He$, the first thought was that the heavy-fermion superconductivity was in a triplet state \cite{cmv-baps},\cite{anderson-dwave}. Soon the expected temperature dependence of various thermodynamic and transport experiments for anisotropic superconductors was calculated and compared with experiments. There were some interesting dilemmas  in understanding the experiments which were partially resolved by realizing that unitarity scattering from impurities \cite{s-rmv-prl86},
\cite{hirschfeld-vollhardt-wofle} produces completely different temperature dependences than perturbative scattering. The temperature dependence of the  ultrasound attenuation could only be understood as due to anisotropic superconductivity with a line of nodes in the gap function. Group theoretical results \cite{blount-volovik-gorkov}
 however told us that a line of nodes is not allowed in the triplet manifold. At this point it was shown that antiferromagnetic fluctuations for which there was a lot of evidence in the heavy-fermion compounds favored \cite{msv-pr86}singlet "d-wave" superconductivity which does indeed  have lines of nodes in three dimensions. This was also the result from a low order analysis of the Hubbard model near its antiferromagnetic transition \cite{scala-loh-hirsch}. There followed a period of great experimental and theoretical work to study the anisotropic superconductivity through various experiments. Much of this work is borrowed in cuprate superconductors because they are indeed also 'd-wave' superconductors, almost certainly due to physics other than antiferromagnetic fluctuations. 
 
 It would be very nice if Josephson type experiments could be done in heavy fermion superconductors so as to pin-down the symmetries in compounds like $UPt_3$, where multiple superconducting transitions are observed. It is interesting to note the possible development of time-reversal violating triplet superconductivity in $Sr_2Ru0_4$ \cite{srruo4}. But this poses its own mystery since there are no ferromagnetic fluctuations, expected to cause triplet pairing, seen by inelastic neutron scattering in this compound.

\section{Outstanding problems}

In recent years  much attention has been devoted to studying the quantum critical point at the antiferromagnetic transition and the competition between superconductivity and antiferromagnetism in heavy fermion compounds. To have a theory of such phenomena, it is first necessary to have a low energy Hamiltonian for the lattice problem.
This has not been worked out yet. This is needed to answer questions like what determines the coherence scale in the Kondo problem, answer questions such as the Exhaustion paradox raised by Nozieres \cite{nozieres-exhaustion}, determine the magnetic and superconducting interaction scales and the relevant operators in the competition between competing instabilities. Systematic hints to answers from some of these questions from experiments \cite{nakafu- pines-fisk} in which one systematically goes from the dilute Kondo limit to the heavy Fermions is very welcome.

To study the critical points, more experimental results for the fluctuation spectra near the critical points are needed. It is interesting that the $\omega/T$ scaling proposed for the quantum criticality of the Cuprates \cite{mfl} is found through a very nice analysis \cite{stockert} of the data in the heavy  in  set of neutron scattering results in $CeCu_{6-x}Au_x$. It appears such spectra appear in a wide variety of conditions which we do not understand.  On the theoretical side, it is doubtless true that the singular undressing of the Kondo effect \cite{maebashi} which must accompany magnetic order as well as the antiferromagnetic criticality are both implicated. There is an interesting set of theoretical ideas proposed related to this \cite{si} which is however tied to two-dimensionality. My guess is that it is not yet the time to declare victory even in the fundamentals of the theory of this problem. 

{\it Acknowledgements}: I learnt an enormous amount on the heavy fermion problems from a long array of experimentalists, including each one of the three fellow speakers (Ott, Steglich and Fisk) at this symposium. I learnt a lot of the physics of magnetism and Kondo effects in the early years through informal lessons from two masters,
P.W. Anderson and C. Herring. My principal and valued collaborators in the heavy fermion problem have been Barbara Jones, Kazumasa Miyake, Stephan Schmitt-Rink, and Yakov Yafet.

\end{document}